\documentclass[12pt]{article}
\usepackage[margin=1in]{geometry}
\usepackage{times}
\usepackage{dcolumn}
\usepackage{float}
\usepackage{graphicx}
\usepackage{siunitx}
\usepackage{dcolumn}
\usepackage{float}
\usepackage{graphicx}
\usepackage{siunitx}
\usepackage{ifpdf}
 \usepackage{latexsym}
 \usepackage{lipsum}
 \usepackage{mathrsfs}
\usepackage{amssymb,amsmath}
\usepackage[FIGTOPCAP]{subfigure}

\def\isundefined#1{TT\fi\ifx#1\undefined}
\if\isundefined\degree
   \newcommand{\degree}{\circ} 
\fi
\if\isundefined\micro
  \newcommand{\micro}{$\mu$} 
\fi

\title{Large conditional single-photon cross-phase modulation}

\author{Kristin M. Beck, Mahdi Hosseini, Yiheng Duan and Vladan Vuleti\'c$^\ast$ \\
\\
\normalsize{Department of Physics and Research Laboratory of Electronics,} \\\normalsize{Massachusetts Institute of Technology, Cambridge, Massachusetts 02139, USA}\\
\\
\normalsize{$^\ast$Corresponding author. E-mail: vuletic@mit.edu}}
\begin{document}

\maketitle
\begin{abstract}
Deterministic optical quantum logic requires a nonlinear quantum process that alters the phase of a quantum optical state by $\pi$ through interaction with only one photon. Here, we demonstrate a large conditional cross-phase modulation between a signal field, stored inside an atomic quantum memory, and a control photon that traverses a high-finesse optical cavity containing the atomic memory. This approach avoids fundamental limitations associated with multimode effects for traveling optical photons. We measure a conditional cross-phase shift of up to $\pi/3$ between the retrieved signal and control photons, and confirm deterministic entanglement between the signal and control modes by extracting a positive concurrence. With a moderate improvement in cavity finesse, our system can reach a coherent phase shift of $\pi$ at low loss, enabling deterministic and universal photonic quantum logic.
\end{abstract}

Universal quantum gates~ \cite{Fredkin:ITP1982, Milburn:PRL1989} can be implemented with an interaction that produces a conditional $\pi$ phase shift by one qubit on another \cite{Chuang:PRA1995}. For photonic qubits, this requires an as-of-yet unrealized strong nonlinear interaction at the single-photon level. Photons do not directly interact with each other, and hence must be interfaced in a medium with a giant nonlinearity while preserving optical coherence~\cite{Schmidt:OL1996, Harris:PRL99}. The strong nonlinearities introduced by interacting Rydberg atoms \cite{Hao:SR2015, Firstenberg:Nature2013,Parigi:PRL2012,Dudin:science2012} and cavity quantum electrodynamic (cQED) systems~\cite{Fushman:Sci:2008,Reiserer:Nature2014,VV:Nature2014} have led to the observation of up to $\pi$ phase shifts between two propagating photons in the same mode. This type of quantum phase switch can be used to sort photons and implement a Bell state analyzer \cite{Witthaut:EPL2012}. The realization of a deterministic and universal optical gate, however, requires cross-phase modulation between distinct optical modes. For light pulses propagating in nonlinear fibers \cite{Kerrinfibre:nph:2010} and nonlinear slow-light media \cite{Hsiang:PRL:XPM:2011,Shiau:PRL:2011}, cross-phase modulation on the order of micro-radians per photon has been observed. In a pioneering cQED experiment two decades ago, Turchette {\it et al.}~measured the average polarization rotation of a weak continuous probe beam by another beam copropagating in the same cavity, and extrapolated a nonlinear phase shift of 0.28~rad per photon \cite{Kimble:PRL1995}. However, the characteristic time of the nonlinearity (the cavity lifetime) in that experiment was much shorter than the photon wavepacket duration necessary to spectrally separate the two modes, which precludes the modulation of the entire wavepacket \cite{Munro:OC2010}. Very recently, a much smaller but conditional cross-phase modulation of 18~\micro rad by a single postselected photon was measured in a nonlinear slow-light system using electromagnetically induced transparency (EIT)  \cite{Feizpour:nphys:2015}. However, as shown by Shapiro  \cite{Shapiro:PRA:2006}, and in an extension to EIT by Gea-Banacloche \cite{Banacloche:PRA:2010}, locality and causality prohibit high-fidelity $\pi$-phase shifting operations between traveling photons. 

\begin{figure*}[!t]
\centerline{\includegraphics[width=.7\columnwidth]{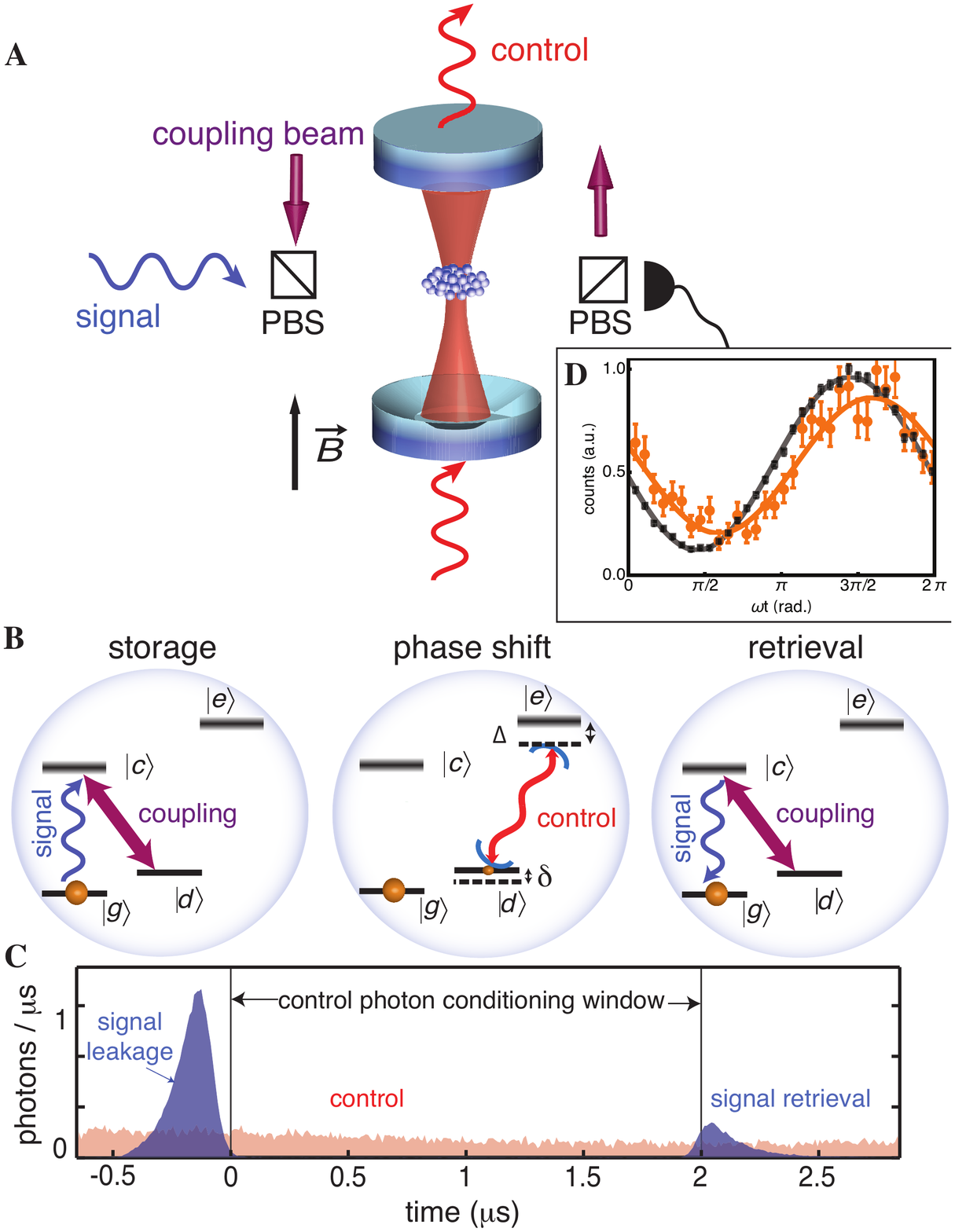}} 
\caption{{\bf Scheme for imprinting large single-photon phase shift onto stored light.} {\bf (A), (B)}  A signal photon traveling orthogonal to the cavity axis is stored as an atomic coherence between states $|g\rangle=|S_{1/2},F=3,m_F=3\rangle$ and $|d\rangle=|S_{1/2},4,4\rangle$ via the EIT process created by coupling light resonant with the $|d\rangle\leftrightarrow|c\rangle=|P_{3/2},3,3\rangle$ transition. A control photon resonant with the optical cavity, and detuned by $\Delta$ from the $|d\rangle$ to $|e\rangle=|P_{3/2},5,5\rangle$ transition, is sent through the cavity during the storage time. The signal photon is retrieved after the control photon leaves the cavity. {\bf (C)} The experimental signal leakage and retrieval (blue), and control (red) light pulses are shown as a function of time. {\bf (D)} The phase of the retrieved signal light is measured without control light (black), and conditioned on the detection of a transmitted control photon (red) by its interference with a co-propagating reference beam (not shown) with $\omega=$30 MHz frequency difference. In this and the following figures, the error bars represent $\pm1$ s.d. of statistical error.}
\label{setup}
\end{figure*}

To realize a giant optical Kerr effect that is not subject to Shapiro's no-go theorem, we coherently store a weak signal pulse in an atomic quantum memory as a collective spin excitation via EIT \cite{Harris:eit97}. A control photon traveling through a high-finesse cavity containing the EIT medium interacts with the entire collective atomic excitation simultaneously, and the stored signal light is retrieved after detecting the control photon (Fig.~\ref{setup}). A similar setup was previously used to implement an optical transistor whose transmission depended on the stored photon number in the quantum memory \cite{chen:science2013}. That work demonstrated that one stored photon can block the transmission of many cavity photons resonant with the atomic transition. The current experiment instead investigates the dispersive regime of atom-cavity coupling: a control photon induces a differential light shift on the two atomic states in the collective excitation, thus shifting the optical phase of the signal light retrieved later. Conversely, a stored signal photon changes the center frequency of the cavity and shifts the phase of a weak control pulse. We measure this cross-Kerr modulation on both signal and control light, conditioned on the detection of a photon in the other mode, while maintaining high fringe visibility. 

Our system consists of an ensemble of laser-cooled ${}^{133}$Cs atoms trapped in a dipole trap tightly focused at the center of a high-finesse optical cavity (Fig.~\ref{setup}A). Initially, the atoms are optically pumped into the state $|g\rangle$. We then make use of the resonant $\Lambda$-type energy-level structure, $|g\rangle\leftrightarrow|c\rangle\leftrightarrow|d\rangle$, to induce EIT. Signal light resonant with the $|g\rangle\leftrightarrow|c\rangle$ transition slowly propagates through the atomic medium while its group velocity is controlled by a strong co-propagating coupling beam resonant with the $|d\rangle\leftrightarrow|c\rangle$ transition (Fig.~\ref{setup}B). By adiabatically switching off this coupling beam (Fig.~\ref{setup}C), the signal photon is stored in the ensemble as an atomic coherence between the $|g\rangle$ and $|d\rangle$ states. In the absence of control photons, we typically store and retrieve more than 10\% of the input signal pulse when we switch on the coupling laser again after 2 \micro s of storage. This retrieval efficiency depends on the ensemble optical depth ({\it OD}) and decoherence rate ($\gamma_0$) of the atomic coherence, measured to be $OD=7$ and $\gamma_0/2\pi=50$~kHz, respectively. 

To measure the conditional phase shift $\phi$ imprinted by one control photon on the stored signal field, a long control pulse (a weak coherent state with less than one photon on average) impinges on the optical cavity during the storage time, and light-shifts the atomic levels. The resulting phase shift of the atomic excitation is mapped onto the signal light upon retrieval, and is measured by comparison with a 30~MHz detuned reference pulse traveling along the signal path. The reference and retrieved signal light mix on a photodetector. The conditional nonlinear phase shift is the difference between the measured signal phase when we detect one transmitted control photon in the conditioning window, compared to the signal phase when no control photon is detected or no control light is applied (Fig.~\ref{setup}D).
 
\begin{figure*}[!t]
\hspace{0 pt} {\bf\large A} \hspace{175 pt} {\bf\large B}\\
 \includegraphics[width=.48\columnwidth]{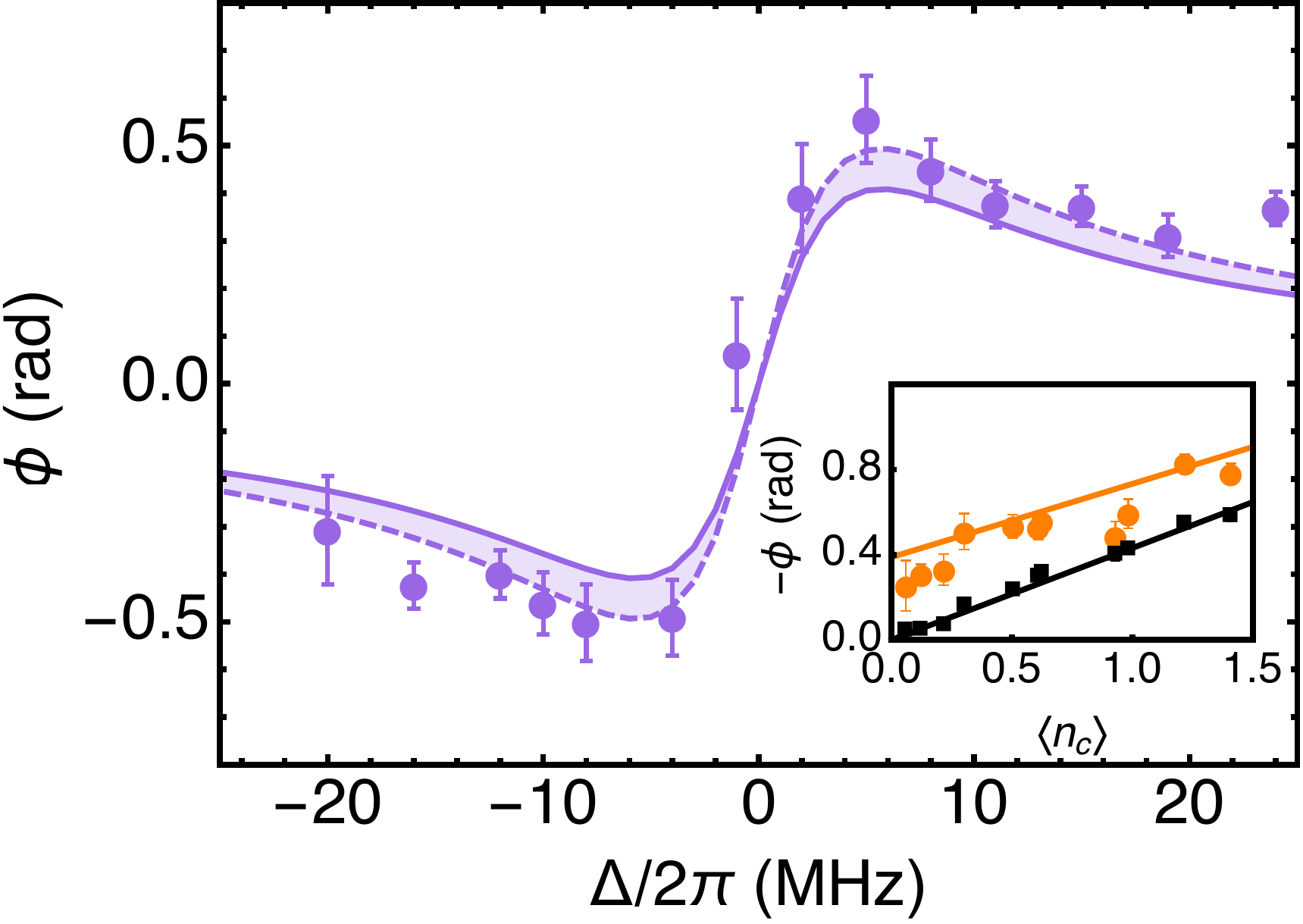} 
 \includegraphics[width=.5\columnwidth]{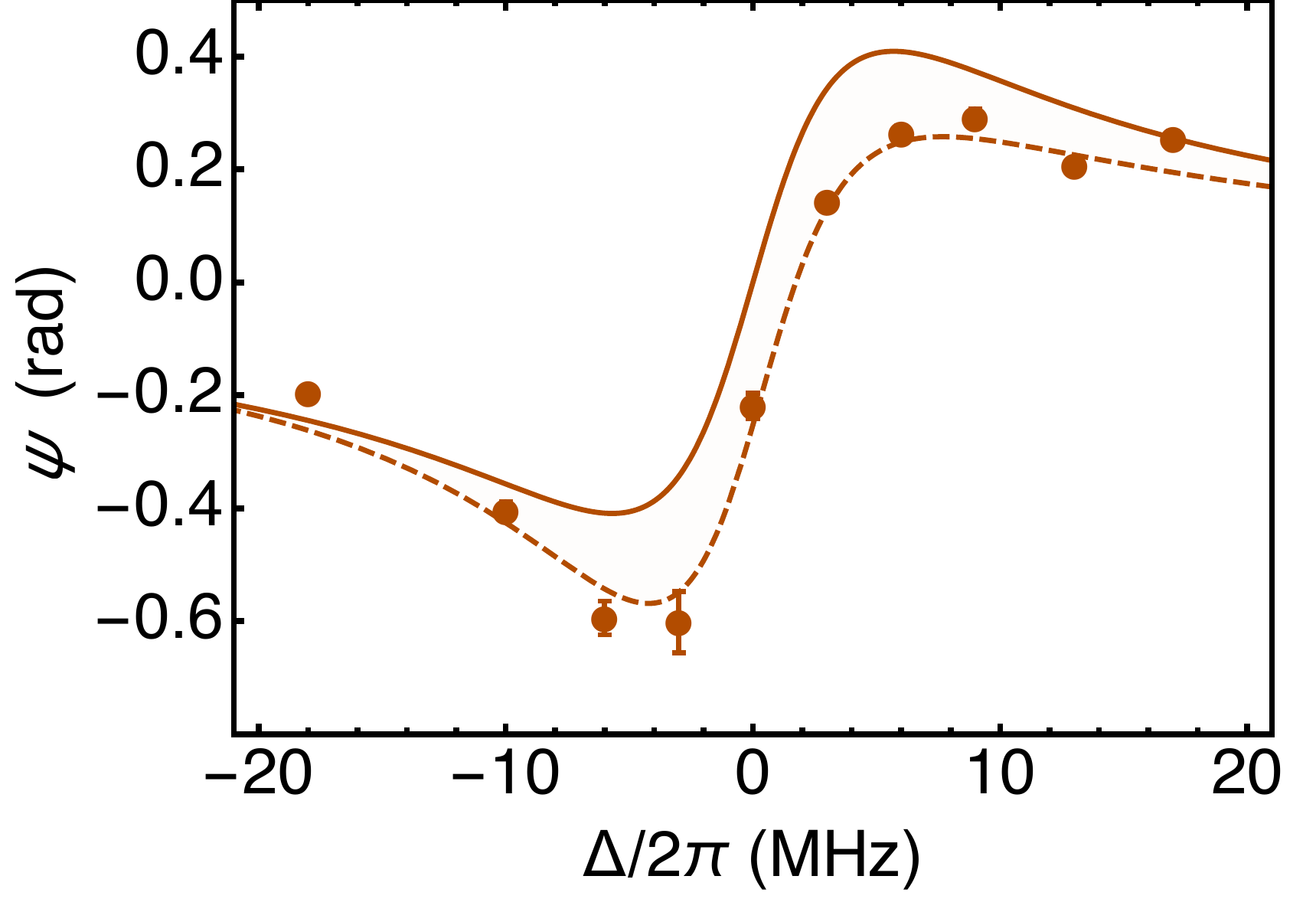}
 \caption{ {\bf Conditional phase shift induced by a single photon.} {\bf (A)}~The phase shift of the stored signal field, conditioned on detecting a control photon, is plotted as a function of control-atom detuning $\Delta$. The solid line is the model prediction for a single control photon in a cavity with cooperativity $\eta=3.8$; the dashed line is this same prediction including corrections for multiple control photons (mean recovered signal photon number $\langle n_s\rangle$=0.3, mean control photon number $\langle n_c\rangle$=0.4). The inset plots the measured average phase shift (black circles) and conditional phase shift (red circles) as a function of $\langle n_c\rangle$ at $\Delta/2\pi=-8$ MHz and $\langle n_s\rangle=0.3$. The average phase shift fits to a line (black) with slope of 0.43(1)~rad/photon that agrees with the expected phase shift per cavity photon of 0.38~rad/photon. The red line is the model's prediction for the conditional phase shift that accounts for contributions from multiple photons. At very low mean control photon numbers, false conditioning on the background counts slightly reduces the measured phase (not included in the model). We also measured the mean phase shift conditioned on detecting no cavity photon (not shown), which is equal to the average phase shift within the error bars. {\bf (B)}~The control phase shift $\psi$, inferred from polarization rotation, conditioned on the detection of a signal photon. The deviation of the experimental data from the theoretical model (solid line) can be explained by a small light-cavity detuning of $\delta_c/2\pi=$25 kHz that is included in the model shown as the dashed line (see Sup. Info.).} 
\label{phaseshift}
\end{figure*}

Fig.~\ref{phaseshift}{\bf A} shows the measured conditional signal phase shift  as a function of the detuning $\Delta$ between the input control light and the atomic transition $|d\rangle\leftrightarrow|e\rangle$. The phase shift results from the light shift $\delta=\eta\frac{ \kappa_0}{2} \text{Re}[\chi]$ of the control photon on the atomic state $|d\rangle$. Here, $\chi=\left(\frac{2\Delta}{\Gamma}+i\right)/\left(1+(\frac{2\Delta}{\Gamma})^2\right)$,  $\eta=4g^2/\kappa_0\Gamma=3.8$ is the spatially averaged cavity cooperativity \cite{Haruka:AAMOP2011}, $\kappa_0=2\pi \times 150$~kHz is the measured empty-cavity linewidth, $2g=2\pi \times 1.6$~MHz is the effective single-photon Rabi frequency, and $\Gamma=2\pi\times5.2$~MHz is the excited-state decay rate. The control-induced nonlinear atomic phase shift is then approximately $\phi=\delta\cdot\tau$, where $\tau=1/\kappa$ is the mean interaction time, and $\kappa = \kappa_0\left(1+\eta\text{Im}[\chi]\right)$ \cite{Haruka:AAMOP2011} is the increased cavity linewidth in presence of a signal photon. We measure a conditional single-photon phase shift of $|\phi|=0.4(1)$ ~rad at $|\Delta|=$8~MHz, in good agreement with the theoretical prediction of $\phi=\frac{1}{2}\eta\text{Re}[\chi]/(1+\eta\text{Im}[\chi])$. The average shift (not conditioned on detecting a control photon) is linear with $\langle n_c\rangle$, the mean input control photon number  in the 2 \micro s conditioning window (inset to Fig.\ref{phaseshift}{\bf B}). The linear slope of 0.43(1)~rad/photon is close to the expected phase shift per cavity photon of 0.39~rad/photon, and the conditional phase shift of 0.4(1) rad measured for $\langle n_c\rangle\ll1$. For $\langle n_c\rangle\gtrsim1$, the contribution from undetected photons increases the measured conditional phase. Throughout this paper, we operate at mean photon numbers $\langle n_c\rangle\le0.5$.

The control light is also affected by the presence of a signal photon: the cavity resonance is shifted by the stored signal light \cite{Marin:PRE2005}, which in turn changes the phase of the transmitted control light. We measure the conditional control phase $\psi$ by using linearly polarized input light on the cavity path, and measuring its polarization change conditioned on detecting one retrieved signal photon. The weakly interacting $\sigma^-$-polarized component thus serves as a phase reference for the strongly interacting $\sigma^+$-polarized control light (see Sup. Info.). This conditional control phase shift $\psi$ is plotted as a function of control-atom detuning in Fig.~\ref{phaseshift}{\bf B}. 

In fact, the combined control-signal optical state can be ideally described as a two-mode entangled state $|\Psi\rangle =p_{00} |0_s0_c \rangle+p_{01}|0_s1_c \rangle+p_{10}|1_s0_c \rangle+p_{11}e^{i\theta}|1_s1_c \rangle$, where $0_s$ ($1_s$) refers to zero (one) signal photon while $0_c$ ($1_c$) represents a $\sigma^-$ ($\sigma^+$)-polarized control photon, $p_{ij}$ is the probability amplitude of being in state $|i_sj_c\rangle$, and $\theta$ is the nonlinear interaction phase. Thus in the ideal system, we expect $\phi=\psi=\theta$. In the presence of decoherence and loss, the two-mode system must be described by a density matrix. We reconstruct the reduced density matrix, $\rho_{ij}~(i,j\epsilon\{0,1\})$, of the outgoing signal and control modes by measuring coincidences between these two paths (see Sup. Info.). We extract a nonlinear phase shift of $\theta=0.45(2)$ rad, and a concurrence \cite{wootters:concurrence01} of $C=0.082(5)>$ 0, after correcting for detection efficiencies and propagation losses. The positive concurrence demonstrates deterministic number-polarization entanglement between the outgoing signal and control light. 

\begin{figure*}
\centerline{ \includegraphics[width=.55\columnwidth]{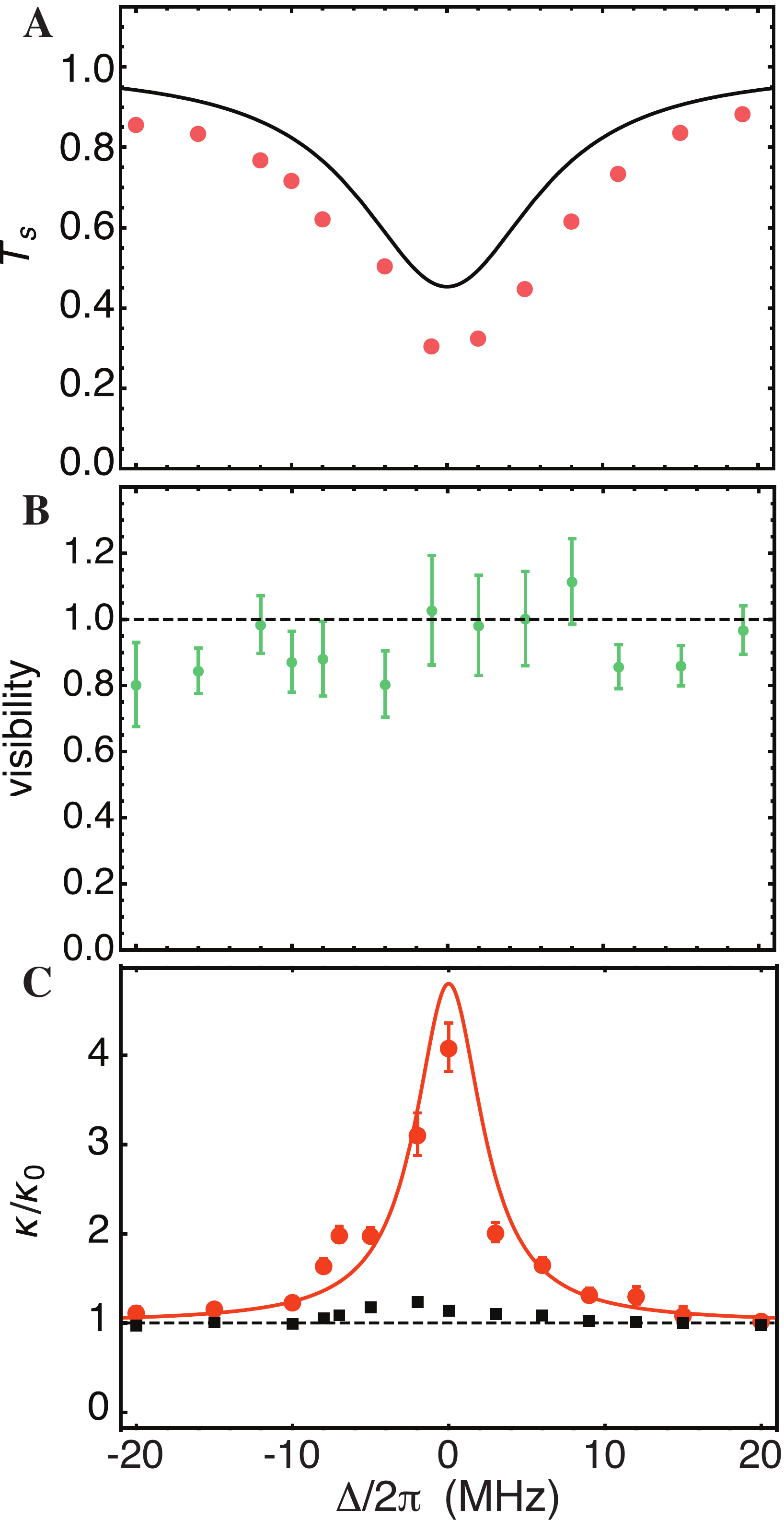}}
\caption{{\bf Signal transmission, signal visibility and cavity linewidth.} {\bf (A)}~The fractional signal transmission conditioned on detecting a control photon, measured at $\langle n_c\rangle=0.4$ and $\langle n_s\rangle=0.3$.  {\bf (B)}~ Fringe visibility of recovered signal light after correction for the signal loss shown in panel (A). The dashed line indicates the ideal visibility. {\bf (C)}~The cavity linewidth conditioned on detecting a signal photon (red circles) and averaged cavity linewidth (black squares), normalized to the bare linewidth $\kappa_0=2\pi \times 150$~kHz, measured for $\langle n_s\rangle= 0.2$. }
\label{vsdeltaac}
\end{figure*}

At a given cooperativity $\eta$, the nonlinear phase shift takes on its maximum value $\phi\simeq\eta/(4\sqrt{1+\eta})$ at a cavity-atom detuning of $\Delta/\Gamma=(1+\eta)/2$, and is accompanied by a slightly reduced signal transmission $T_s/T_0=e^{-\eta/2(1+\eta)}=0.67$ for $\eta=3.8$.  The latter ($T_s/T_0$) is due to the scattering of photons into free space that destroys the collective spin excitation associated with the stored signal photon. Fig.~\ref{vsdeltaac}{\bf A} shows this signal recovery efficiency conditioned on the detection of a control photon. The solid curve is the theoretical expectation for transmission taking into account the signal loss due to the scattering of the control photon, given by $T_s/T_0=\exp(-\eta\text{Im}[\chi]\kappa_0/\kappa)$. Additional losses are responsible for the remaining small deviation between the experimental data and this curve (see Sup. Info.). 

As there is uncertainly on the time scale $\kappa^{-1}$ when a control photon enters or exit the cavity, we expect a randomization $\delta\phi$ of the nonlinear phase  \cite{Munro:OC2010} at the level of $\delta\phi/\phi=(\kappa\tau_p)^{-1}\sim0.25$, where $\tau_p=2$~\micro s is the input control pulse length. This would limit the visibility of the recovered phase to about 0.99 at  $\phi=0.4$ rad. The visibility of our phase beatnote after correcting for the transmission loss is shown in Fig.~\ref{vsdeltaac}{\bf B}. This measurement yields an average visibility of 0.9(1) at $\Delta=-8$ MHz that is consistent with expected visibility reduction and appears to be independent of $\Delta$.

The lifetime of the cavity photon, $1/\kappa$, decreases in the presence of the atomic excitation (stored signal photon) that can scatter light out of the cavity. To confirm this aspect of our model, we excite the cavity with a short pulse (200 ns) and measure the cavity decay time conditioned on detecting a stored signal photon. In Fig.~\ref{vsdeltaac}{\bf C}, we plot the conditional cavity linewidth $\kappa$ as a function of the control-atom detuning $\Delta$. A single atom in state $|d\rangle$ increases the cavity linewidth by $\kappa/\kappa_0 = 1+\eta\text{Im}[\chi]$ \cite{Haruka:AAMOP2011}, which is plotted as the theoretical curve in Fig.~\ref{vsdeltaac}A. The observed increase of the cavity linewidth agrees with the theory. Remarkably, the cavity lifetime is shortened even in those instances when the signal photon is detected, i.e. the scattering of the cavity photon into free space did not actually occur. 

\begin{figure*}[!t]
\centerline{\includegraphics[width=.5\columnwidth]{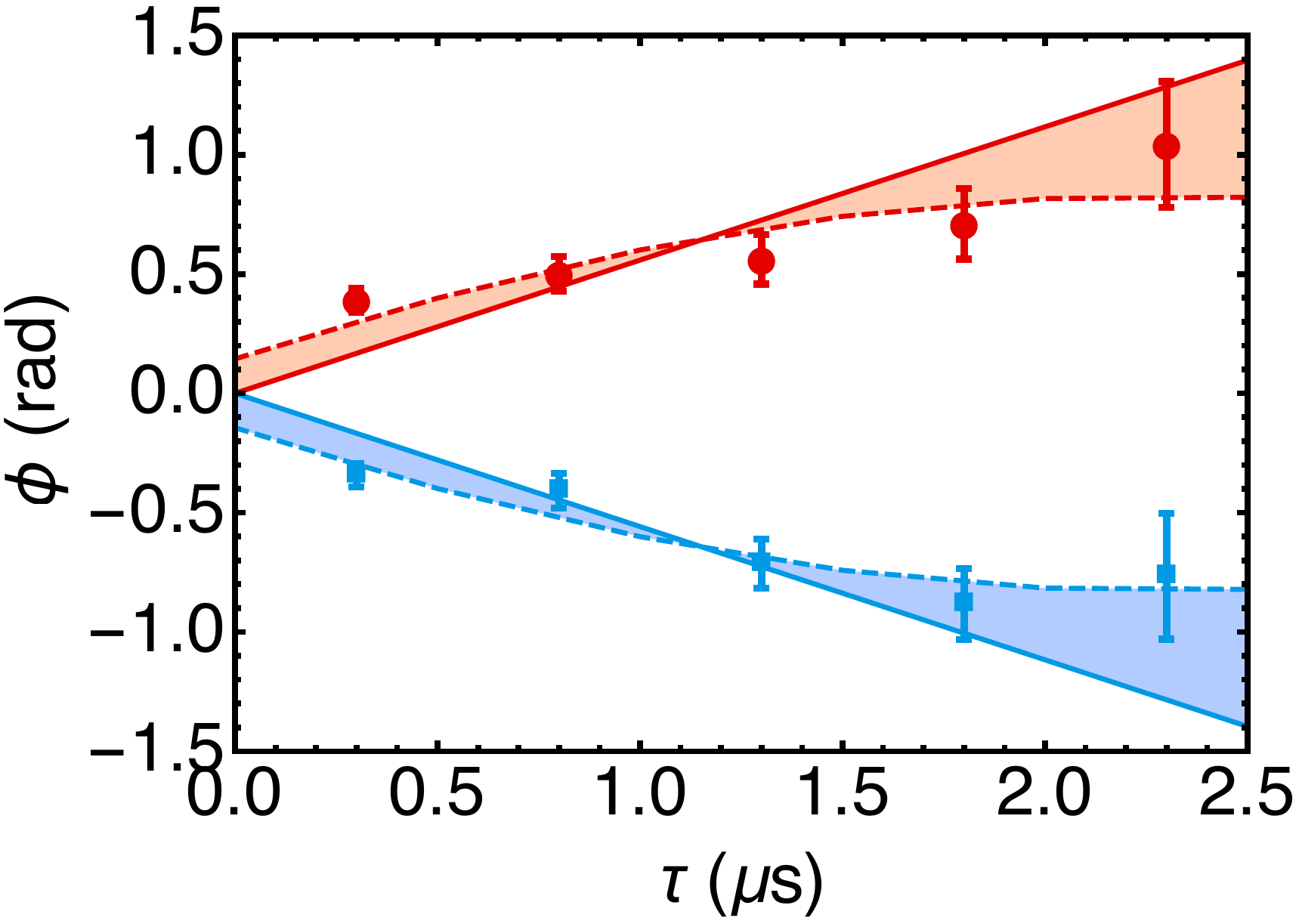}}
\caption{{\bf Signal phase shift conditioned on control photon dwell time.} The conditional phase shift for $\Delta/2\pi=\pm8$~MHz is plotted as a function of conditioning time $\tau$ for signal pulses stored for 3 \micro s, $\langle n_c\rangle=0.8$ and $\langle n_s\rangle=0.6$. The conditioning window is 0.5~\micro s. The solid lines model the phase as linear in control photon dwell time; the dashed lines are the complete model predictions and include the effect of multiple control photons (important at early conditioning times) and background counts (important at long conditioning times) (see Sup. Info.).}
\label{vstime}
\end{figure*}

In this short-pulse excitation ($\tau_p\ll\kappa^{-1}$) limit, we can directly measure the change in the imprinted phase shift with control photon dwell time. The imprinted phase shift on the signal light should be proportional to the time the control photon spends in the cavity exerting a light shift on the spin wave. Therefore, we can increase the phase shift by postselecting on control photons that exit the cavity later than average. In Fig.~\ref{vstime}, we plot the resulting phase shift as a function of the conditioning time for control-atom detunings $\Delta/2\pi=\pm 8$ MHz. The observed conditional phase shift increases for long control photon dwell times. The largest phase shift we observe is 1.0(4)~rad, 2.5 times larger than the phase shift observed for long pulses with $\tau_p>\kappa^{-1}$. 

In conclusion, we have measured a conditional phase shift of 0.4(1)~rad onto a weak coherent state by a single photon using quasi-monochromatic light, and up to 1.0(4)~rad by using a short control pulse and postselecting on photons that remain in the system for longer than average. The underlying interaction entangles the outgoing signal and cavity modes as verified by a positive concurrence. Using a single-sided cavity would enable us to reach phase shifts of $\pi$, on atomic resonance, as was recently measured between a photon and an atom  \cite{Reiserer:Nature2014}. Alternatively, increasing the cavity cooperativity in the present geomentry enables phase shifts approaching $\pi$ detuned from resonance.  Such large and efficient conditional phase modulation at the single photon level would enable deterministic optical quantum logic~\cite{agrwala:pra2005}, the engineering of cluster states  \cite{kim:OC2015,Gao:PRL2010}, and entanglement concentration  \cite{Filip:PRA2003}.  

The authors would like to thank M. Lukin and J. Thompson for enlighting discussions. This work was supported by the NSF, and MURI grants through AFOSR and ARO. K.M.B. acknowledges support from NSF IGERT under grant 0801525.

\bibliography{Ref}
\bibliographystyle{pnas}

\clearpage


\clearpage 
\section*{Supplemental Material}
\makeatletter \renewcommand{\thefigure}{S\@arabic\c@figure} \renewcommand{\thetable}{S\@arabic\c@table} 
\setcounter{figure}{0}

 \makeatother
\makeatletter

\subsection*{Experimental Details}

The Cs atoms in our experiment are held in a far off-resonant dipole trap that is focused at the cavity waist. This trap is fromed by 32~mW of 937~nm  light focused through an in-vacuum lens to give an expected transverse waist of $2.5$~\micro m at the atoms. The corresponding calculated trap frequencies are $\omega_{\text{radial}}/2\pi=46$~kHz and $\omega_{\text{axial}}/2\pi= 4$~kHz. From absorption images of the atomic cloud, we measure the atoms to have a transverse rms radius of 5(1)~\micro m and an axial rms radius of 19(1)~\micro m. 

The atom-cavity coupling $g$, and thus the cooperativity $\eta=4g^2/\kappa\Gamma$, varies along the standing wave of the cavity axis and with the radial extent of the cavity mode. The extent of the atomic cloud and its placement determine the effective cooperativity we realize in the experiment. The maximal cooperativity $\eta_0=\frac{24\mathcal{F}/\pi}{k^2w_c^2}=8.6(1)$ is determined by the wavevector $k=2\pi/\lambda$ where $\lambda=852.347$~nm, the cavity waist $w_c=35.5(2)$~\micro m, and the cavity finesse $\mathcal{F}=77.1(5)\times10^3$. The maximal value is realized on the cavity axis at an antinode of the cavity standing wave. The effective cooperativity is the value averaged over all possible atomic positions
\begin{eqnarray}
\eta=\eta_0\iiint\rho(x,y,z)\cos^2(kz)e^{-\frac{x^2+y^2}{2 w_c^2}} dx\ dy\ dz
\end{eqnarray}
where $\rho(x,y,z)$ is the normalized atomic density. From the above atomic distribution, we predict an effective cooperativity $\eta=3.8(1)$. This number is used to plot the expected theory curves in all of the figures in the main text.

{\bf Signal transmission.} While the primary recovery loss is due to interactions with the control photon, two other factors reduce the recovery. Since the atomic cloud extends beyond the cavity waist, there is recovery loss due the inhomogeneous phase shift imprinted by the cavity photon, a result of the spatially inhomogeneous cavity coupling. The occasional presence of more than one cavity photons reduces further the signal photon's survival probability beyond the expected conditional transmission (shown in Fig.3 (A)). \\

{\bf Input control photon number.} In order to calculate the input control photon number $\langle n_c\rangle$ we measure the mean photon number transmitted through the cavity and divide it by the detection efficiency (0.45), fiber collection efficiency (0.7), cavity out-coupling efficiency (0.66), and atom-induced cavity transmission \cite{Haruka:AAMOP2011} 
\begin{eqnarray}
T_c=\frac{1}{(1+\langle n_s\rangle\eta \text{Im}[\chi])^2+(\frac{2\delta_c}{\kappa}+\langle n_s\rangle\eta \text{Re}[\chi])^2}
\end{eqnarray}
taking into account the mean stored signal photon number $\langle n_s\rangle<1$. The calculated $\langle n_c\rangle$ is equivalent to the mean input control photon number at the input of a fully impedance matched cavity during the 2~\micro s storage time.
   
\subsection*{Conditional phase shift from a coherent state}

The equally weighted total conditional signal phase shift is given by
\begin{eqnarray}
\phi (t)= \text{Arg}\left[\sum_{m}P(m|n)e^{m\phi}\right]
\end{eqnarray}
where $\phi$ is the phase shift of a single cavity photon induced on the signal light and $P(m|n)$ is probability of having $m$ photons conditioning on detecting $n$ photons given by \cite{Feizpour:nphys:2015} 

\begin{eqnarray}
&&P(m|n_c)\\ \nonumber &&=  \left( {\begin{array}{cc} m+n_{bg}   \\  n  \ \end{array} } \right) \epsilon_d^{n} (1-\epsilon_d)^{n_{bg}+m-n} P(m)
\end{eqnarray}
where the background counts $n_{bg}=R_b t$ with detected background rate $R_b$ and conditioning time window $t$, $\epsilon_d$ is the detection efficiency of the conditioning path, and $P(m)$ is the probability for $m$ photons to be observed in a given coherent state.

\subsection*{Conditional cross-phase modulation}

In order to calculate the conditional phase shift we diagonalize the system Hamiltonian approximately given by 

\begin{eqnarray}
\frac{\hat{H}}{\hbar} &=& \omega_a \hat{S} +\omega_c \hat{c}^{\dag}\hat{c} + \Omega \hat{c}^{\dag}\hat{c} \hat{S}+\int{d\omega~\omega(\hat{b}_{\omega}^{\dag}\hat{b}_{\omega}+\hat{d}_{\omega}^{\dag}\hat{d}_{\omega})}
\\ \nonumber
&&+ i\sqrt{\frac{\kappa_0}{4\pi}}\int{d\omega \{ [\hat{b}_{\omega}^{\dag}\hat{c} - \hat{c}^{\dag} \hat{b}_{\omega} ] +[\hat{d}_{\omega}^{\dag}\hat{c} - \hat{c}^{\dag} \hat{d}_{\omega}] \} }
\end{eqnarray}
where $\omega_a$ and $\omega_c$ are frequencies of the atomic spin ($\hat{S}=|d\rangle\langle d|-|g\rangle\langle g|$) and intra-cavity optical field ($\hat{c}$), respectively. By adiabatically eliminating the excited state, we define the effective atom-light coupling strength, $\Omega=g^2\Delta/(\Delta^2+(\Gamma/2)^2)$, with one-photon detuning $\Delta$, single-photon Rabi frequency $2g$,  and excited state decay rate of $\Gamma$. The last two terms of the Hamiltonian account for energy and multimode coupling of input (or reflected) and transmitted light represented by modes $\hat{b}_{\omega}$ and $\hat{d}_{\omega}$, respectively. To diagonalize the Hamiltonian we follow Ref.\cite{Leroux:PRA2012} and define the following operators
\begin{eqnarray}
\hat{a}_{\omega} &=& \frac{1}{(\omega-\omega_c-\Omega \hat{S})^2+(\kappa_0/2)^2}\left[i\sqrt{\frac{\kappa_0}{2\pi}}\hat{c}^{\dag} +\right.\\ \nonumber
&&\frac{\kappa_0}{2\sqrt{2\pi}}\hat{c}^{\dag}\int{\hat{d}\omega' \frac{\hat{b}_\omega^{\dag}+\hat{d}_{\omega}^{\dag}}{\omega-\omega'}}+\\ \nonumber
&& \left. \frac{1}{\sqrt{2}}(\omega-\omega_c-\Omega \hat{S})(\hat{b}_{\omega}^{\dag}+\hat{d}_{\omega}^{\dag})\right]\\
\hat{\bar{a}}_{\omega} &=& \frac{1}{\sqrt{2}} (-\hat{b}_{\omega}^{\dag}+\hat{d}_{\omega}^{\dag})
\end{eqnarray}
to rewrite the Hamiltonian as
\begin{eqnarray}
\frac{\hat{H}}{\hbar} &=& \omega_a \hat{S} + \int{d\omega~\omega(\hat{a}_{\omega}^{\dag}\hat{a}_{\omega}+\hat{\bar{a}}_{\omega}^{\dag}\hat{\bar{a}}_{\omega})}
\end{eqnarray}

Note that these operators have the following commutation relation:
\begin{eqnarray}
{[} \hat{a}_{\omega} , \hat{a}_{\omega'}^{\dag} {]} &=& \delta (\omega-\omega') \\
{[} \hat{\bar{a}}_{\omega} , \hat{\bar{a}}_{\omega'} ^{\dag} {]} &=& \delta (\omega-\omega')\\
{[}\hat{ a}_{\omega} , \hat{\bar{a}}_{\omega'} ^{\dag} {]} &=& 0.
\end{eqnarray}
For transmitted cavity light, the final state of the system after an interaction time of $t$ 
\begin{eqnarray}
|\Psi \rangle &=& e^{-i (m\Phi+\omega_a t \hat{S})} |\Psi_a\rangle \otimes \\ \nonumber
&& \frac{1}{\sqrt{m!}}\bigg[\int{d\omega B(\omega) \hat{d}_{\omega}^{\dag}}\bigg]^{m}|0\rangle
\label{eq:phi}
\end{eqnarray}
where $|\Psi_a\rangle$ and $|0\rangle$ are field eigenstates, $B(\omega)$ is the pulse amplitude spectrum and $m$ is the photon number. The phase
 $\Phi=\arctan(\frac{2\delta_c}{\kappa}+\phi)$ and 
\begin{eqnarray}
\phi& =& \frac{\eta}{2} \text{Re}[\chi]\frac{\kappa_0}{\kappa}\\
\kappa&=&\kappa_0(1+\text{Im}[\chi])\\
\chi&=&\left(\frac{2\Delta}{\Gamma}+i\right)/\left(1+(\frac{2\Delta}{\Gamma})^2\right).
\end{eqnarray}
Here $\kappa_0=2\pi \times 150$~kHz is the measured empty-cavity linewidth, $2g=2\pi \times 1.6$~MHz is the single-photon Rabi frequency, and $\Gamma=2\pi\times5.2$~MHz is the excited state decay rate. Using Eq.\ref{eq:phi}, the conditional phase shift between a single cavity photon and one stored atomic excitation can then be written

\begin{eqnarray}
\Phi= \arctan\left(\frac{2\delta_c}{\kappa}+\frac{\eta}{2} \frac{\text{Re}[\chi] }{1+\text{Im}[\chi]}\right) -\arctan\left(\frac{2\delta_c}{\kappa_0}\right) 
\label{eq:arctan}
\end{eqnarray}
where $\delta_c=\omega-\omega_c$ is detuning of light from empty cavity resonance. In the limit of small detuning $\delta_c/\kappa\ll1$, the conditional phase shift is approximately given by 
\begin{eqnarray}
\Phi\simeq\phi=\frac{\eta}{2} \frac{\text{Re}[\chi] }{1+\text{Im}[\chi]}.
\end{eqnarray}

\subsection*{Conditional control phase shift}
In the main text, we focus on the phase shift measurement of the stored atomic excitation due to its interaction with a control photon. The control light also undergoes a phase shift. This phase shift is the result of shift in cavity resonance frequency imposed by atoms in the cavity mode.  To measure the phase shift on the control light we linearly polarize the input control light and measure its conditional polarization rotation. 

\begin{figure*}
	\centerline{\includegraphics[width=.55\columnwidth]{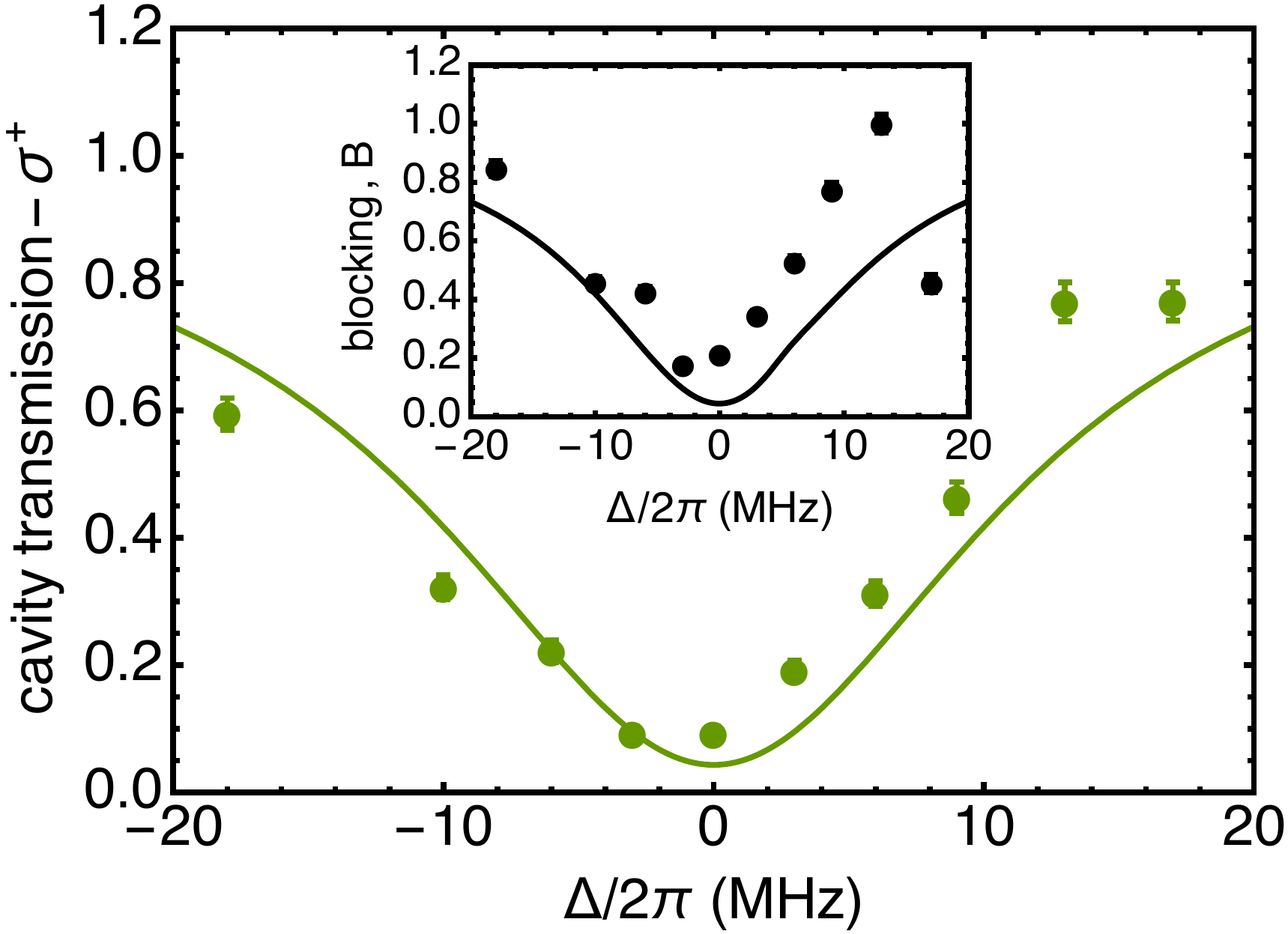}}
	\caption{Cavity transmission for right-circularly polarized light conditioned on detecting one retrieved signal photons is plotted as a function of light-atom detuning. The solid line represents the theoretical expectation. Inset shows the blocking factor, $B$, of the linearly polarized control light plotted conditioned on detecting a signal photon. The solid line represents the theoretical expectation. The deviation of the data from the theory in the inset can be explained by the interaction of $\sigma^-$-polarized light with excited states other than $|F'=5,m_f=3\rangle$ not taken into consideration in the model. }
\label{blocking}
\end{figure*}

The effective cooperativity of the $\sigma^-$-polarized light is reduced by a factor of 45 compared to $\sigma^+$-polarized light and detuned by 8 MHz from $|F=4,m_f=4\rangle\to |F'=5, m_f=3\rangle$ transition when $\sigma^+$ light is resonant with  $|F=4,m_f=4\rangle\to |F'=5, m_f=5\rangle$ transition. The interaction of linearly polarized light then predominantly the interaction of the $\sigma^+$ polarization component. Therefore, we can measure the phase shift on $\sigma^+$ light as a polarization rotation on the outgoing control light. This polarization rotation is measured using two photon counters placed at the two ports of a polarizing beam splitter (PBS) after the cavity. By rotating the polarization before the PBS by 45 $^{\degree}$ and subtracting the two photon-count rates, $d_1$ and $d_2$, a signal proportional to $\sin(\psi)$ is obtained as $\frac{d_1-d_2}{d_1+d_2}=\frac{2\sqrt{B}}{(1+B)}\sin(\psi)$. Here, $B$ is the blocking factor that accounts for different transmission of $\sigma^+$ and $\sigma^-$ components of light. In presence of a stored signal photon, the transmission on cavity resonance for $\sigma^+$-polarized light is reduced by this factor $B\simeq(1+\eta \text{Im}[\chi])^2+(\eta \text{Re}[\chi])^2$ compared to the $\sigma^-$ transmission. We separately measure $B$ (inset of Fig.\ref{blocking}) by detecting the cavity transmission for each circular polarization conditioned on retrieving a signal photon. This measured blocking factor allows us to extract the cavity phase shift (Fig.\ref{blocking}). We attribute the asymmetry in the shape of the phase shift as a function of detuning plotted in Fig.3{\bf D} to non-zero detuning of light from the cavity. The solid line is the theoretical expectation (Eq.\ref{eq:arctan}) assuming a light-cavity detuning $\delta_c=0~$kHz.  As the sign of the frequency shift in the cavity resonance conditioned on detecting one recovered signal photon changes with the sign of light-atom detuning $\Delta$, having a non-zero $\delta_c$ results in total shift of the cavity resonance that is different in magnitude for positive and negative detuning $\Delta$. This causes an asymmetry in the phase shift as a function of $\Delta$ (Fig.2(B)). This change with light-atom detuning can be understood as a different interaction time: the lifetime of cavity photons effectively reduces with detuning of light from the cavity resonance.

\subsection*{Reconstruction of density matrix}

Based on the density matrix reconstruction provided by James, {\it et al.}\cite{James:PRA2001} for polarization-entangled photons, we developed a method to reconstruct the full density matrix of a number-polarization entangled state $|\Psi \rangle$ by measuring coincidences in the $|0_s\rangle$ and $|1_s\rangle$ signal photon basis and the $|0_c\rangle$ and $|1_c\rangle$ control polarization basis, where for simplicity we use 0 and 1 representation for zero and one signal photon and also different circularly polarized control photons, {\it i.e.} $0_c \leftrightarrow\sigma^-$ and $1_c \leftrightarrow\sigma^+$. Coherently interfering the signal state with a phase reference allows us to project the signal state into an arbitrary superposition $|0_s\rangle+e^{i \theta_s}|1_s\rangle$. To project the control light into the desired state, $|0_c\rangle+e^{i \theta_c}|1_c\rangle$, we use a half-wave plate (HWP) and quarter-wave plate (QWP) after the cavity followed by a PBS. The phases $\theta_s$ and $\theta_c$ can be chosen for each measurement by changing phase of the signal reference light and polarization of the control light after the cavity, respectively. However, the auxilliary optical phase reference adds a complication: as this reference light is not part of the signal photon state, we need to normalize out its contribution to the measured coincidences to ensure the reconstructed density matrix is independent of the reference light intensity.
\subsubsection*{Coincidence measurements}
In total, 16 coincidence measurements are required to reconstruct the complete density matrix. To arrive at the derived coincidences for the output  number-polarization state, we measure the raw coincidences, normalize and then reconstruct coincidences for $|\Psi \rangle$.

 {\bf Measure raw coincidences.} The tomographic states equivalent to Ref.\cite{James:PRA2001} for coincidences $n_{\nu}~ (\nu=1,2,..16)$, are listed for signal and control modes in Table S1.

In our experiment, these tomographic states are obtained through interference measurements of the signal with an optical phase reference and polarization measurement of the control light. For the signal path, the phase reference light is a frequency shifted beam copropagating with the signal. This light mixes on the detector to form a beatnote with a period of about 33~ns. All relative phase angles $\theta_s$ are measured in a single dataset. For the control path, we linearly polarize the input cavity light so that the phase between control light (the $\sigma^+$-polarized component) and its reference ($\sigma^-$-polarized component) appears as a polarization rotation at the cavity output. We use a HWP, QWP and a polarizing beam splitter to analyze the output at different projection angles $\theta_c$.

We measure the raw coincidences $n_{\nu}~(\nu=1,2,..16)$ on a pair of single photon counters. Measuring in four configurations we measure all 16 tomographic states:

\begin{itemize}
\item \underline{Coincidences for $\nu=1-4$} are measured without signal phase reference while control light is measured in $\sigma^+$ or $\sigma^-$ polarization. The coincidences are then extracted from these measurements. For example, $n_3$ is the number of times that a signal photon (without reference) and a $\sigma^+$-polarized control photon are simultaneously detected, and $n_1$ is number of coincidences where no signal photon (without reference) and a $\sigma^-$-polarized control photon are detected.

\item \underline{Coincidences for $\nu=5-8$} are measured with a phase reference light in the signal mode. To reconstruct interference fringes, signal photons are conditioned on the detection of a $\sigma^-$ or $\sigma^+$ control photon. The resulting coincidence counts form a beatnote. The coincidence counts $n_\nu$ are then the number of coincidences at the phase $\theta_s$, which we extract from a fit to the counts at all phases.  Detector counts on the signal path at $\theta_s=0\ (3\pi/2)$ corresponds to projecting the $|\Psi \rangle$ to $|0_s\rangle+|1_s\rangle$ ($|0_s\rangle-i|1_s\rangle$). To be concrete, $n_6$ is the number of coincidences with a $\sigma^+$ photons detected on the control at $\theta_s=3\pi/2$ and corresponds to a tomographic measurement onto signal state $|0_s\rangle-i|1_s\rangle$ and control state $|1_c\rangle$. 

\item \underline{Coincidences for $\nu=12-15$} are measured with no signal reference while the control light is projected into different superposition state of $\sigma^-$ and $\sigma^+$-polarized light. The HWP and QWP placed after the cavity and before a PBS followed by a single photon detector, sets the measurement basis and thus the relative phase angle $\theta_c$. The coincidence count $n_{14}$, for example, is the number of times 1 photon is detected in the signal path and 1 photon is detected on the control path with the analysis HWP at $\pi/8$, which corresponds to a relative phase $\theta_c=\pi/2$ between $\sigma^-$ and $\sigma^+$ polarized light. This is a tomographic measurement onto the signal state $|1_s\rangle$ and control state $|0_c\rangle +i|1_c\rangle$.

\item \underline{Coincidences for $\nu=9-11$ and $\nu=16$} are measured with signal phase reference while control light is projected into different superposition state of $\sigma^-$ and $\sigma^+$-polarized light.  These elements are determined as for $\nu=5-8$: detected signal photons are conditioned on detecting 1 control photon (now at a relative phase angle $\theta_c$), and the coincidence counts $n_\nu$ are the number of coincidences at phase $\theta_s$. 

 \end{itemize}
 
 {\bf Normalizing coincidences.} The coincidences for $\nu=5,6,...,16$ are measured using signal phase reference that needs to be normalized out. To do this, we calculate the interference parameter $\mathcal{I}_\nu$ which takes on values between -1 and 1. On the signal path, the interference parameter is just $\vartheta\cos{\theta}$, that is the value of a zero-centered interference fringe with contrast $\vartheta$ and a phase difference of $\theta$.

To obtain $\mathcal{I}_{\nu}$, we make two additional measurements. We measure the signal and signal phase reference beatnote with input light on the signal path only (no control light), as well as the linearly-polarized control without signal light at different HWP angles to reconstruct a complete fringe. This measures the contrast in the absence of interactions. We then calculate $\mathcal{I}$ by:
\begin{enumerate}
\item We subtract the averaged value of the raw coincidences (coincidence number averaged over all signal phases or angles of the HWP in the control path) from $n_\nu$ ($\nu=5,...,16$).
\item We divide the coincidence number by the total number of detected counts in the conditioning port.
\item We finally divide through by the fringe amplitude measured in our additional measurements without interaction to correct for non-unity contrast without interactions, for example due to power imbalance between the signal and phase reference.
\end{enumerate}
 
  {\bf Reconstruct coincidences.} The values of $\mathcal{I}_{\nu}$ together with $n_{\nu}~ ({\nu=1,2,3,4})$ are used to reconstruct the coincidence numbers for $|\Psi\rangle$ alone for all 16 tomographic measurements. We first rescale coincidences $n_{\nu}~(\nu=1,2,3,4)$ to correct for detection losses, $\epsilon_d$. The efficiency $\epsilon_d\simeq0.2$ for both signal and control modes. Projecting the general output state $|\Psi\rangle= p_{00}|0_s0_c\rangle+p_{01} e^{i\phi_{01}}|0_s1_c\rangle+p_{10} e^{i\phi_{10}}|1_s0_c\rangle+p_{11} e^{i\phi_{11}}|1_s1_c\rangle$ (with probability amplitudes $p_{ij}$ and phase shift $\phi_{ij}$) to the relevant tomographic state for each $\nu=5,...,16$ allows us to evaluate coincidences in terms of $\mathcal{I}$ and $n_{\nu}~ ({\nu=1,2,3,4})$. For example, to reconstruct $n_5$, we project $|\Psi\rangle$ onto $(|0_s\rangle+e^{i\theta_s}|1_s\rangle)|0_c\rangle$. The expectation value for this parameter is then 
   \begin{eqnarray}
  \langle{n}_5\rangle&=&| \langle\psi| \hat{a}_s^{\dag}\hat{a}_c^{\dag}\hat{a}_c\hat{a}_s|\Psi\rangle|^2\nonumber\\
  &=& p_{00}^2+p_{10}^2+2p_{00}p_{10}\vartheta_{5}\cos (\theta_s+\phi_{10})
   \end{eqnarray}
  where $\vartheta_{5}$ is the corresponding contrast, $\hat{a}_s$ ($\hat{a}^\dagger_s$) and $\hat{a}_c$ ($\hat{a}^\dagger_c$) are signal and control lowering (raising) operators, respectively, and the relative phase angle $\theta_s=3\pi/2$. Repeating this for each $n_{\nu} (\nu=5,...,16)$, and expressing total coincidences in terms of $\mathcal{I}$ and $n_{\nu}~ ({\nu=1,2,3,4})$ we find:
 \begin{eqnarray}
n_5&=& \frac{n_1+n_4}{2}+\sqrt{n_1 n_4} \mathcal{I}_{5} 
 \end{eqnarray} 
 \begin{eqnarray}
n_6&=& \frac{n_2+n_3}{2}+\sqrt{n_2 n_3} \mathcal{I}_{6}
 \end{eqnarray} 
 \begin{eqnarray}
 n_6&=&\frac{n_2+n_3}{2}+\sqrt{n_2 n_3} \mathcal{I}_{7} 
  \end{eqnarray} 
 \begin{eqnarray}
n_8&=&\frac{n_1+n_4}{2}+\sqrt{n_1 n_4} \mathcal{I}_{8}
 \end{eqnarray} 
 \begin{eqnarray}
n_9&=&\frac{n_1+n_2+n_3+n_4}{4} + \\ \nonumber 
&& \frac{1}{2} \sqrt{(\sqrt{n_1n_4}+\sqrt{n_2n_3})^2+(\sqrt{n_2n_4}-\sqrt{n_1n_3})^2} \mathcal{I}_{9}
 \end{eqnarray} 
 \begin{eqnarray}
n_{10}&=&\frac{n_1+n_2+n_3+n_4+2\sqrt{n_1n_2}+2\sqrt{n_3n_4}}{4}    \\ \nonumber
&&+  \frac{1}{2} (\sqrt{n_1}+\sqrt{n_2}) (\sqrt{n_3}+\sqrt{n_4})\mathcal{I}_{10} 
 \end{eqnarray}
 \begin{eqnarray}
n_{11}&=&\frac{n_1+n_2+n_3+n_4+2\sqrt{n_1n_2}+2\sqrt{n_3n_4}}{4}\\ \nonumber
&& +\frac{1}{2} (\sqrt{n_1}+\sqrt{n_2}) (\sqrt{n_3}+\sqrt{n_4})\mathcal{I}_{11}
 \end{eqnarray} 
 \begin{eqnarray}
 n_{12}&=& \frac{n_1+n_2}{2}+\sqrt{n_1 n_2} \mathcal{I}_{12}
  \end{eqnarray} 
 \begin{eqnarray}
 n_{13}&=& \frac{n_3+n_4}{2}+\sqrt{n_3 n_4} \mathcal{I}_{13} 
  \end{eqnarray} 
 \begin{eqnarray}
n_{14}&=&\frac{n_3+n_4}{2}+\sqrt{n_3 n_4} \mathcal{I}_{14}
 \end{eqnarray} 
 \begin{eqnarray}
n_{15}&=&\frac{n_1+n_2}{2}+\sqrt{n_1 n_2} \mathcal{I}_{15} 
 \end{eqnarray} 
 \begin{eqnarray}
n_{16}&=&\frac{n_1+n_2+n_3+n_4}{4} +\\ \nonumber
&&\frac{1}{2} \sqrt{(\sqrt{n_1n_4}+\sqrt{n_2n_3})^2+(\sqrt{n_2n_4}-\sqrt{n_1n_3})^2} \mathcal{I}_{16} 
\end{eqnarray}
\subsubsection*{Experimental density matrix}

The final coincidences can then be used to reconstruct the experimental density matrix following Ref.\cite{James:PRA2001}:
 \begin{eqnarray}
\hat{\rho}_{ex} = \frac{\sum_{\nu=1}^{16} {M_{\nu} n_{\nu}} }{\sum_{\nu=1}^{4} {n_\nu}}
\end{eqnarray}
where matrices $M_{\nu}$ in these bases are provided below. By doing so, we arrive at the following measured experimental density matrix:
\newline
\newline
\[
\hat{\rho}_{ex} = \left( {\begin{array}{cccc}    
0.6358 & 0.4319 - 0.07635 i & 0.1337 - 0.00026 i &  0.00154 - 0.0222 i\\
 0.4319 + 0.07635 i & 0.3205 & 0.1292 - 0.07199 i &  0.0593 - 0.01282 i\\
 0.1337 + 0.00026 i & 0.1292 + 0.07199 i & 0.02899 &  0.0184 - 0.0084 i\\
 0.00154 + 0.0222i & 0.0593 + 0.01282 i&  0.0184 + 0.0084 i& 0.0146\end{array} }\right)
\]
\subsubsection*{Reconstruction of the physical density matrix}

In order to reconstruct the physical density matrix that most probably describes the measurement results we use the Maximum Likelihood (Maxlik) method as outlined in Ref.\cite{James:PRA2001}. This requires finding minimum of the following function:
 \begin{eqnarray}
&&\mathcal{L}(t_1,t_2,...,t_{16})\\ \nonumber &&=\sum_{\nu=1}^{16}{\frac{(\mathcal{N}\langle \psi_{\nu} | \hat{\rho}_p(t_1,t_2,...,t_{16})| \psi_{\nu}\rangle - n_{\nu})^2}{2 \mathcal{N} \sigma_{\nu}^2}}
\end{eqnarray}
where 
\begin{eqnarray}
\hat{\rho}_p(t) &=&\frac{\hat{T}^{\dag} \hat{T}}{\text{Tr}[\hat{T}^{\dag} \hat{T}]},\\
\hat{T} &=& \left( {\begin{array}{cccc} t_1 & 0 & 0 &0 \\ t_5+i t_6 & t_2 & 0 & 0 \\ t_{11}+i t_{12} & t_7+i t_8 & t_3 & 0\\
t_{15}+i t_{16} & t_{13}+i t_{14} & t_9+i t_{10} & t_4   \ \end{array} } \right)\\
\mathcal{N}&=&\sum_{n=1}^4{n_{\nu}}
\end{eqnarray}
and $\sigma_{\nu}$ is the standard deviation for the $\nu$th coincidence measurement given approximately by $\sqrt{n_{\nu}}$ (Poisson noise). The initial estimation of $t_1,... t_{16}$ is obtained using the inverse relationship by which elements of $\hat{T}$ can be expressed in terms of elements of $\hat{\rho}_{exp}$ (see Ref.\cite{James:PRA2001}). After the Maxlik reconstruction of the density matrix and subtracting the global phases, we obtain the following matrix

\[
\hat{\rho}_{p} =  \left( {\begin{array}{cccc} 
0.6315 & 0.4174 & 0.1375 & 0.0495 - 0.0239i\\
0.4174& 0.321224 & 0.0996- 0.0035i &0.0527 - 0.0248i\\
 0.1375 & 0.0996 + 0.0035i & 0.0319&0.0153- 0.0054 i\\
  0.0495+ 0.0239i & 0.0527+ 0.0248i & 0.0153 + 0.0054i & 0.0154\end{array} }\right)
\]
where we calculate Tr$[\rho^2]=0.92$ as a measure of purity. The concurrence is evaluated as $C(\rho)= \text{max}(0,\lambda_1-\lambda_2 -\lambda_3 -\lambda_4)$, where $\lambda_i$'s are the square roots of the eigenvalues of $\rho\rho'$ in descending order, $\rho'= (\sigma_y\otimes\sigma_y)\rho^*(\sigma_y\otimes\sigma_y)$, and $\sigma_y$ is Pauli $y$ matrix. We obtain a concurrence of 0.082(5) and nonlinear phase shift measured as $\phi_{nl}= \text{Arg}(\rho_p[1,4])=  0.45(2)$ that agrees with the measured conditional phase shift. To estimate the error in determining concurrence and phase, we randomly sample half of the data 100 times and reconstruct the density matrix each time and find the standard deviation of the concurrence and phase calculations. Note that the maximum concurrence at this phase shift using the same coherent states as in our experiment is 0.11. In the case where input states are equal superposition of $|0_s\rangle$ and $|1_s\rangle$, a concurrence on the order of $|\sin (\phi/2)|$ is ideally achievable \cite{Sanders:JOSAB:2010}. 

\subsubsection*{The $M_{\nu}$ matrices}
The $M_{\nu}$ matrices defined below are from Ref.\cite{James:PRA2001} with corrected typos in $M_{2}$ and  $M_{14}$.
 \begin{eqnarray}
  M_1 &=&\frac{1}{2} \left( {\begin{array}{cccc}       2      &  -(1-i)  &  -(1+i)  &  1 \\
                                                                                   -(1+i)  &       0   &      i       & 0  \\
                                                                                   -(1- i)  &       -i   &      0       & 0  \\
                                                                                          1  &       0   &      0       & 0   \\ \end{array} } \right),\nonumber
                                                                                           \end{eqnarray} 
 \begin{eqnarray}
M_2 &=&\frac{1}{2} \left( {\begin{array}{cccc}       0      &  -(1-i)  &       0  &  1 \\
                                                                                   -(1+i)  &       2   &      i       &  -(1+i) \\
                                                                                          0   &      -i   &      0       & 0  \\
                                                                                          1  &    -(1- i) &      0       & 0   \\ \end{array} } \right),\nonumber
                                                                                           \end{eqnarray} 
 \begin{eqnarray}
 M_3 &=&\frac{1}{2} \left( {\begin{array}{cccc}       0      &  0 &       0  &  1 \\
                                                                                      0      &   0   &      i       &  -(1+i) \\
                                                                                          0   &      -i   &      0       &  -(1-i)   \\
                                                                                          1  &    -(1- i) & -(1+i) & 2   \\ \end{array} } \right),   \nonumber
                                                                                           \end{eqnarray}
 \begin{eqnarray}
 M_4 &=&\frac{1}{2} \left( {\begin{array}{cccc}       0      &  0 &      -(1+i) &  1 \\
                                                                                      0      &   0   &      i       &  0 \\
                                                                                    -(1-i)   &      -i   &      2       &  -(1-i)   \\
                                                                                          1  &    0    & -(1+i) & 0   \\ \end{array} } \right)  ,\nonumber
                                                                                           \end{eqnarray} 
 \begin{eqnarray}
 M_5 &=&\frac{1}{2} \left( {\begin{array}{cccc}       0      &  0 &      2i &  -(1+i) \\
                                                                                      0      &   0   &      (1-i)     &  0 \\
                                                                                    -2i   &      (1+i)   &      0       &  0   \\
                                                                                        -(1-i)  &    0    & 0  & 0   \\ \end{array} } \right)   ,\nonumber
                                                                                         \end{eqnarray} 
 \begin{eqnarray}                                                                                   
 M_6 &=&\frac{1}{2} \left( {\begin{array}{cccc}       0      &  0 &      0  &  -(1+i) \\
                                                                                      0      &   0   &      (1-i)     &  2i \\
                                                                                     0   &      (1+i)   &      0       &  0   \\
                                                                                        -(1-i)  &    -2i    & 0  & 0   \\ \end{array} } \right) ,\nonumber
                                                                                         \end{eqnarray} 
 \begin{eqnarray}        
      M_7 &=&\frac{1}{2} \left( {\begin{array}{cccc}       0      &  0 &      0  &  -(1+i) \\
                                                                                      0      &   0   &      -(1-i)     &  2 \\
                                                                                     0   &      -(1+i)   &      0       &  0   \\
                                                                                        -(1-i)  &    2    & 0  & 0   \\ \end{array} } \right)   ,  \nonumber
                                                                                         \end{eqnarray}
 \begin{eqnarray}
       M_8 &=&\frac{1}{2} \left( {\begin{array}{cccc}       0      &  0 &      2  &  -(1+i) \\
                                                                                      0      &   0   &      -(1-i)     &  0 \\
                                                                                     2  &      -(1+i)   &      0       &  0   \\
                                                                                 -(1-i)  &    0    & 0  & 0   \\ \end{array} } \right)    ,\nonumber
                                                                                  \end{eqnarray} 
 \begin{eqnarray}         
  M_9 &=&\frac{1}{2} \left( {\begin{array}{cccc}       0      &  0 &      0  &  i \\
                                                                                      0      &   0   &      -i     &  0 \\
                                                                                     0   &      i   &      0       &  0   \\
                                                                                        -i  &    0    & 0  & 0   \\ \end{array} } \right)   ,\nonumber
                                                                                         \end{eqnarray} 
 \begin{eqnarray}             
   M_{10} &=&\frac{1}{2} \left( {\begin{array}{cccc}       0      &  0 &      0  &  1 \\
                                                                                      0      &   0   &      1     &  0 \\
                                                                                     0   &      1   &      0       &  0   \\
                                                                                        1  &    0    & 0  & 0   \\ \end{array} } \right)   ,\nonumber
                                                                                         \end{eqnarray} 
 \begin{eqnarray}             
      M_{11} &=&\frac{1}{2} \left( {\begin{array}{cccc}       0      &  0 &      0  &  i \\
                                                                                      0      &   0   &      i     &  0 \\
                                                                                     0   &      -i   &      0       &  0   \\
                                                                                        -i  &    0    & 0  & 0   \\ \end{array} } \right)   ,\nonumber
                                                                                         \end{eqnarray} 
 \begin{eqnarray}
     M_{12} &=&\frac{1}{2} \left( {\begin{array}{cccc}       0      &  2 &     0  &  -(1+i) \\
                                                                                          2      &   0   &      -(1+ i)     &  0 \\
                                                                                          0    &   -(1- i)   &      0       &  0   \\
                                                                                     -(1-i)  &    0    & 0  & 0   \\ \end{array} } \right)  ,\nonumber
                                                                                      \end{eqnarray} 
 \begin{eqnarray}   
     M_{13} &=&\frac{1}{2} \left( {\begin{array}{cccc}       0      &  0 &     0  &  -(1+i) \\
                                                                                            0      &   0   &      -(1+ i)     &  0 \\
                                                                                          0    &   -(1- i)   &      0       &  2   \\
                                                                                     -(1-i)  &    0    & 2  & 0   \\ \end{array} } \right)   ,\nonumber
                                                                                      \end{eqnarray} 
 \begin{eqnarray}
       M_{14} &=&\frac{1}{2} \left( {\begin{array}{cccc}       0      &  0 &     0  &  -(1- i) \\
                                                                                            0      &   0   &      (1- i)     &  0 \\
                                                                                          0    &   (1+ i)   &      0       &  -2i   \\
                                                                                     -(1+i)  &    0    & 2i  & 0   \\ \end{array} } \right)  ,\nonumber
                                                                                      \end{eqnarray} 
 \begin{eqnarray}      
     M_{15} &=&\frac{1}{2} \left( {\begin{array}{cccc}       0      &  -2i &     0  &  -(1- i) \\
                                                                                            2i       &   0   &      (1- i)     &  0 \\
                                                                                          0    &   (1+ i)    &      0       &  0   \\
                                                                                     -(1+i)  &    0       &   0     & 0   \\ \end{array} } \right)  ,\nonumber
                                                                                      \end{eqnarray} 
 \begin{eqnarray}  
        M_{16} &=&\frac{1}{2} \left( {\begin{array}{cccc}       0      &  0 &      0  &  1 \\
                                                                                      0      &   0   &      -1     &  0 \\
                                                                                     0   &      -1   &      0       &  0   \\
                                                                                        1  &    0    & 0  & 0   \\ \end{array} } \right) .         
     \end{eqnarray}

\clearpage

 \begin{table}
\begin{center}
\begin{tabular}{ c  c c  c  c  c c}
\hline
$\nu$     & &   signal  & control  &  $\theta_s$  & $\theta_h$ & $\theta_q$ \\
\hline
1    & &  $|0_s\rangle$  &  $|0_c\rangle$  &  NA  &  0 & $\pi/4$\\
2    & &  $|0_s\rangle$  &  $|1_c\rangle$  &  NA  &  0 & -$\pi/4$\\
3    & &  $|1_s\rangle$  &  $|1_c\rangle$  &  NA  &  0 & -$\pi/4$\\
4    & &  $|1_s\rangle$  &  $|0_c\rangle$  &  NA  &  0 & $\pi/4$\\
5    & &  $|0_s\rangle-i|1_s\rangle$  &  $|0_c\rangle$  &  $3\pi/2$  &  0 & $\pi/4$\\
6    & &  $|0_s\rangle-i|1_s\rangle$  &  $|1_c\rangle$  &  $3\pi/2$  &  0 & -$\pi/4$\\
7    & &  $|0_s\rangle+|1_s\rangle$  &  $|1_c\rangle$  &  0  &  0 & -$\pi/4$\\
8    & &  $|0_s\rangle+|1_s\rangle$  &  $|0_c\rangle$  &  0 &  0 & $\pi/4$\\
9    & &  $|0_s\rangle+|1_s\rangle$  &  $|0_c\rangle-i|1_c\rangle$  &  0  &  -$\pi/8$ & 0\\
10    & &  $|0_s\rangle+|1_s\rangle$  &  $|0_c\rangle+|1_c\rangle$  &  0  &   0& 0\\
11    & &  $|0_s\rangle-i|1_s\rangle$  &  $|0_c\rangle+|1_c\rangle$  &  $3\pi/2$  &   0 & 0\\
12    & &  $|0_s\rangle$  &  $|0_c\rangle+|1_c\rangle$  & NA  &   0 & 0\\
13    & &  $|1_s\rangle$  &  $|0_c\rangle+|1_c\rangle$  & NA  &   0 & 0\\
14    & &  $|1_s\rangle$  &  $|0_c\rangle+i|1_c\rangle$  & NA  &   $\pi/8$ & 0\\
15    & &  $|0_s\rangle$  &  $|0_c\rangle +i |1_c\rangle$  & NA  &   $\pi/8$ & 0\\
16    & &  $|0_s\rangle-i|1_s\rangle$  &  $|0_c\rangle +i |1_c\rangle$  & $3\pi/2$  &  $\pi/8$ & 0\\
\hline
\end{tabular}
\end{center}
\caption{The 16 measurements needed to reconstruct the density matrix in the photon-number basis of signal mode and polarization-basis of the control mode. The measurement phase angle is listed as NA when there is no phase reference on the signal path.  $\theta_h$ and $\theta_q$ represent angles of the HWP and QWP placed after the cavity.}
\label{default}
\end{table}     

\clearpage                                                                             
                   
\end{document}